\documentclass[a4paper]{easychair} % A4 is needed for the abstract book

\usepackage[UKenglish]{babel} 
\newcommand\fl{\mathbin{\rightarrow}}
\newcommand\ttt{\mathsf{t}}
\newcommand\eee{\mathsf{e}}
\newcommand\systF{\mathsf{F}}

\newcommand\ma[1]{\textit{``#1''}}

\title{Type-theoretical natural language semantics:\\ on the system $\systF$  for meaning assembly}
\titlerunning{On the system $\systF$  for meaning assembly}
\author{
Christian Retor{\'e}\inst{1}\thanks{Funded by ANR projects Loci and Polymnie.}
}
\institute{
{\'E}quipe \textsc{Melodi},  IRIT , UPS, 118 route de Narbonne, 31062 Toulouse Cedex 9\\ 
  \textit{(\& LaBRI, Universit{\'e} de Bordeaux, 351 cours de la Lib{\'e}ration, 33405 Talence cedex)}\\
}
\authorrunning{Ch. Retor{\'e}}

\newcommand\donotcirculate[1]
{\noindent\raisebox{#1}[0ex]{\makebox[0ex][l]{\hspace*{0.0\textwidth} 
\tt{
\hspace*{5cm} In TYPES 2013 Workshop (Toulouse, April 2013) 
}}}}

\begin{document}
\maketitle
\donotcirculate{40ex}
Roughly speaking, the compositional semantics analysis of natural language consists in mapping 
a sentence to a logical formula which depicts its meaning. To do so, a lexicon endow 
each word with a partial formula (a $\lambda$-term over the base type $\ttt$ for propositions and $\eee$ for individuals), 
and the (binary) parse tree of the sentence specifies for each node 
which subtree is the function and which one is the argument. 
Hence a $\lambda$-term corresponding to the whole parse tree, 
and by reduction one obtains a $\lambda$-term of type $\ttt$ which corresponds to a formula of higher order logic.
This classical process which uses Church's representation of formulae as simply typed $\lambda$-terms is the basis of the so-called Montague semantics, see e.g. \cite[Chapter 3]{MootRetore2012lcg}. 

But it is more accurate to have many individual base types rather than just $\eee$. This way, the 
application of a predicate to an argument only happens when it makes sense. 
For instance sentences like \ma{The chair barks.} or \ma{Their five is fast.} are easily ruled out when there are several types for individuals 
by saying that \ma{barks} and \ma{fast}  respectively 
require arguments of type \ma{dog} and \emph{physical or animated object}. 
Nevertheless, such a type system needs to incorporate some flexibility. Indeed, in the context of a football match, the second sentence makes sense, because \ma{their five} may be understood as a player. 

Accounts of these meaning transfers received a lot of attention since the 80's and 
some formal accounts were proposed in particular by Asher \cite{asher-webofwords}.  
We also proposed an account using the 
system 
$\systF$ of Girard (1971) 
\cite{Girard2011blindspot}
and we studied the compositional properties of such a system in particular for quantifiers, plurals and generic elements, as well as 
the related lexical issues like meaning transfers, copredication, fictive motion,...   
see e.g. \cite{BMRjolli,Retore2012rlv}. 
Our system works as follows: the lexicon provides each word with a main $\lambda$-term, the "usual one" which specifies the argument structure of the word, by using refined types (e.g. "sleeps: $\lambda x^{ani} \underline{sleeps}(x)$" requires an animated subject). In addition, the lexicon may endow each word with a finite number of $\lambda$-terms (possibly none) that implement meaning transfers. For instance a \ma{book} may be turned into a physical object $\phi$ or into an informational content $I$, by constants of respective types $book\fl I$ and $book\fl \phi$ (sometimes the $\lambda$-terms are more complex than simple constants). That way a sentence like \ma{This book is heavy but interesting.} can be properly analysed. Some meaning transfers  are declared to be \emph{rigid} in the lexicon: rigidity prohibits the use of other meaning transfers thus we can block  \ma{* Liverpool defeated Chelsea and decided to build new docks.} 
because the meaning transfer from a town to a football club is declared in the lexicon as rigid. 

In such a setting it is very convenient to quantify over types, for instance quantifiers $\forall, \exists$ may be given a type $ \Lambda \alpha. (\alpha \fl \ttt) \fl \ttt$
It also allows a factorised treatment of conjunction: each time an object $x$ of type $\xi$ can be viewed both as an object 
of type $\alpha$ (via an optional term $f_0:\xi\fl\alpha$) to which the property $P$ applies and as an object of type $\beta$  
(via an optional term $g_0:\xi\fl\beta$)
to which the property $Q$ applies, one can express that $x$ enjoys $P\land Q$. This polymorphic \ma{and} simply is: $\Lambda \alpha \Lambda \beta
\lambda P^{\alpha \fl \ttt} \lambda Q^{\beta\fl \ttt} 
 \Lambda \xi \lambda x^\xi 
 \lambda f^{\xi\fl\alpha} \lambda g^{\xi\fl\beta}.\ 
%\linebreak \hspace*{1ex} \hfill 
%\\ &&\hspace*{3em}
(\land^{\ttt\fl\ttt\fl\ttt} \ (P \ (f \ x)) (Q \ (g \  x))) 
%\end{array}
$. 
To sum up the logical system, we also have  two layers  but they slightly vary from the ones of Montague semantics. 
Our \emph{meta logic} (a.k.a. glue logic) is system $\systF$  with many base types $\ttt$, $\eee_i$ (instead of simply typed $\lambda$-calculus with $\ttt$, and $\eee$)  
Our \emph{logic for semantic representations} is many-sorted higher-order logic ($\eee_i$ instead of a single sort $\eee$). 
For representing quantification, we actually prefer to use Hilbert's $\epsilon$ and $\tau$-terms constructed  with two constants $\epsilon,\tau: \Lambda \alpha.\ (\alpha \fl \ttt) \fl \alpha$ and one for generic elements \cite{Retore2012rlv}. 
An important but rather easy property holds: if the constants define an $n$-order $q$-sorted logic, any ($\eta$-long) normal $\lambda$-term of type $\ttt$ does actually correspond to a formula of $n$-order $q$-sorted logic (possibly $n=\omega$). 

We preferred system $\systF$  to modern type theories (MTT) used by Luo 
\cite{Luo2011lacl} or to the categorical logic of Asher \cite{asher-webofwords}
because of its formal simplicity and absence of variants --- as the terms are issued from the lexicon by means of syntactic rules,  
$\systF$  terms with a problematic complexity are avoided. 
There are two properties of Luo's approach \cite{Luo2011lacl}  that would be welcome: a proper
 notion of subtyping, mathematically safe and linguistically relevant, and predefined inductive types with specific reduction rules. 
Subtyping  is most welcome in particular to represent ontological inclusion (a \ma{human being} is an \ma{animal}, thus predicates that apply to \ma{animals} also apply to \ma{human beings}). Coercive subtyping as developed by Luo and Soloviev\cite{Soloviev00coercioncompletion} sounds promising for $\systF$ (and other notions as well \cite{citeulike:1367405}). The key property of coercive subtyping is that there is at most one subtyping map between any two complex types, provided that there is at most one subtyping map between any two base types. 
Predefined types, inductive types would be most welcome in our setting,  e.g. integers as in G{\"o}del's system T and finite sets of $\alpha$-objects.  Of course, $\systF$  can encode such types, but such encodings are far from natural. Reduction in such a setting is related to the work of Soloviev and Chemouil \cite{Soloviev2003} 
%  
% But there are at least two features of type theory that Luo used for natural language semantics 
% that we would like to add to system $\systF$.  
%
%
%Predefined types, e.g. for finite subsets, for integers, and other elementary base types are also welcome. 
%
%As one can imagine, 
%
%Hence we propose to show how one can adds these two features to system $\systF$ : 
%\begin{description} 
%\item[predefined types]  
%\item[subtyping]  How can we safely add  
%If relevant other subtyping notions  for system $\systF$  can also be considered like \cite{citeulike:1367405}
%\end{description} 
The key point is to show that normalisation and confluence are preserved and that there is no constant-free and closed term of a false type. We shall also illustrate the linguistic relevance of these extensions, which are already included in Moot's semantical and semantical parser for French. \cite{moot10grail}
% create the bibliography
\bibliography{bigbiblio}   % refers to MyBib.bib
\end{document}